\documentclass[aps,prb,twocolumn,floatfix,superscriptaddress]{revtex4-1}

\usepackage{graphicx}
\usepackage{bbm}
\usepackage{braket}
\usepackage{amsmath}
\usepackage{amsfonts}
\usepackage{graphicx}
\usepackage{amssymb}
\usepackage{color}
\usepackage{hyperref}
\usepackage{listings}
\usepackage{array}

\def\sss{\scriptscriptstyle\rm}

\newcommand{\un}[0]{\underline{n}}

\newcommand{\dE}[0]{_{\sss E, N}}
\newcommand{\dD}[0]{_{\sss \Delta, N}}
\newcommand{\de}[0]{_{\sss \varepsilon, N}}

\newcommand{\dN}[0]{_{\sss N}}
\newcommand{\drNr}[0]{_{\sss N}^r}

\newcommand{\dNa}[0]{_{\sss {N+1}}}
\newcommand{\daNa}[0]{_{\sss N}^a}

\newcommand{\dNr}[0]{_{\sss {N-1}}}

\newcommand{\dNHs}[0]{_{\sss  { H {\sigma},N  } } }
\newcommand{\dNH}[0]{_{\sss  { H,N  } } }

\newcommand{\dNL}[0]{_{\sss  {L,N}}}

\newcommand{\dNLs}[0]{_{\sss  { L {\sigma},N  } } }

\newcommand{\dNaH}[0]{_{\sss {H,N+1}}}

\newcommand{\dNaHs}[0]{_{\sss {H {\sigma},N+1  }}}

\newcommand{\dNpd}[0]{_{\sss {N+\delta N}}}
\newcommand{\dNmd}[0]{_{\sss {N-\delta N}}}

\newcommand{\dks}[0]{_{\sss KS}}
\newcommand{\dksN}[0]{_{\sss KS,N}}

\def\dkss{_{\sss KS \sigma}}

\def\dzs{_{\sss 0 {\sigma}}}

\def\ds{_{\sss \sigma}}

\def\dSLs{_{\sss Sl \sigma}}

\newcommand{\dHartree}[0]{_{\sss H}}

\newcommand{\xc}[0]{ {\sss{XC}} }

\newcommand{\dxc}[0]{_{\sss{XC}} }
\newcommand{\dxcN}[0]{_{\sss{XC, N}} }

\newcommand{\dx}[0]{_{\sss{X}} }
\newcommand{\dxN}[0]{_{\sss{X, N}} }

\def\dxs{_{{\sss X} \sigma}}
\def\dxcs{_{{\sss XC} \sigma}}
\def\dxjs{_{{\sss X j} \sigma}}

\def\dNHs{_{{\sss  H {\sigma}, N} }}

\def\djs{_{{\sss j} \sigma}}
\def\dis{_{{\sss i} \sigma}}

\newcommand{\uKLI}[0]{^{\sss {KLI}}}

\newcommand{\uLDA}[0]{^{\sss {LDA}}}

\newcommand{\uxLDA}[0]{^{\sss{X,LDA}}}

\def\uD{^{2}}

\newcommand{\br}[0]{{\bf r}}
\renewcommand{\vr} {{\bf r}}

\newcommand{\be}{\begin{equation}}
\newcommand{\beq}{\begin{equation}}
\newcommand{\ee}{\end{equation}}
\newcommand{\bea}{\begin{eqnarray}}
\newcommand{\eea}{\end{eqnarray}}
\newcommand{\ben}{\begin{equation}}
\newcommand{\een}{\end{equation}}
\newcommand{\ba}{\begin{array}}
\newcommand{\ea}{\end{array}}

\newcommand{\nn}{\nonumber}

\newcolumntype{C}[1]{>{\centering\let\newline\\\arraybackslash\hspace{0pt}}m{#1}}

\usepackage[dvipsnames]{xcolor}

\begin{document}

\title{Fundamental gaps of quantum dots on the cheap}
\author{Alberto Guandalini}
\affiliation{Dipartimento di Scienze Fisiche, Informatiche e Matematiche, Universit{\`a} di Modena e Reggio Emilia, Via Campi 213A, I-41125 Modena, Italy}
\email{alberto.guandalini@unimore.it}
\affiliation{CNR -- Istituto Nanoscienze, Via Campi 213A, I-41125 Modena, Italy}
\author{Carlo A. Rozzi}
\affiliation{CNR -- Istituto Nanoscienze, Via Campi 213A, I-41125 Modena, Italy}
\author{Esa R{\"a}s{\"a}nen}
\affiliation{Laboratory of Physics, Tampere University of Technology, FI-33101 Tampere, Finland}
\author{Stefano Pittalis}
\affiliation{CNR -- Istituto Nanoscienze, Via Campi 213A, I-41125 Modena, Italy}
\email{stefano.pittalis@nano.cnr.it}

\begin{abstract}
We show that the fundamental gaps of quantum dots can be accurately estimated at the computational effort of a standard ground-state calculation supplemented with a {\em non} self-consistent step of negligible cost, all performed within density-functional theory at the level of the local-density approximation.
\end{abstract}

\maketitle

\sloppy

\section{Introduction}\label{I}

In single-electron transport through a semiconductor quantum dot~\cite{RM02} (QD), 
an electron can pass from one reservoir (the source) to another (the drain) when a voltage is applied. 
In this process, an electron is first added to and then removed from the dot. Assuming a weak-coupling of the dots to the reservoirs, 
the addition of an electron requires to overcome the so-called charging energy.
Coulomb blockade oscillations  arise in the conductance from the sequence of charging and discharging 
the QD.~\cite{MWL91} The interval between neighboring Coulomb peaks is the difference between the 
removal energy $E\drNr = E\dNr - E\dN$  and (the negative of) the addition energy  $E\daNa =  E\dN - E\dNa$, 
where $E\dN$ is the  ground state energy of the QD with $N$ electrons. 
Thus, the {\em fundamental gap} is defined as
\begin{align}\label{FG1}
G\dE:=&  E\drNr - E\daNa \nn \\=&  E\dNr-2E\dN+E\dNa. 
\end{align}
This quantity is useful in the evaluation of the electronic properties of a QD, especially in the context of applying them in a circuit or in lattices such as QD cellular automata.

In Kohn-Sham (KS) density-functional theory (DFT)  \cite{dreizler1990density,parr1994density,fiolhais2008primer} --
 through the ionization potential theorem~\cite{PPLB82,PhysRevLett.51.1884,PhysRevA.30.2745,PhysRevA.29.2322,PhysRevB.31.3231,PL97} --
the fundamental gap can also be expressed 
 as follows~\cite{SS85}
\begin{equation}\label{FG2}
G\de  = \varepsilon\dNaH-\varepsilon\dNH\
\end{equation}
where 
$\varepsilon\dNH$ is the energy of the highest (H) occupied  KS level for the system with $N$ electrons -- hence the subscript $N$;
the corresponding orbital may be referred to as the highest occupied ``molecular" orbital (HOMO). Note that, throughout this work, we are primarily concerned with non-degenerate levels.

By mixing states with different integer electron numbers and, thus,
switching from DFT to Ensemble-DFT (EDFT), 
 one finds
that the fundamental gap can be expressed in terms of two contributions~\cite{PPLB82,PL97}
\begin{equation}\label{FG3}
G\dD = \Delta\dksN+\Delta\dxcN \ ,
\end{equation} 
 where
 \begin{equation}\label{FGKS}
\Delta\dksN =\varepsilon\dNL-\varepsilon\dNH 
\end{equation}
is the energy gap between the last occupied  and the first unoccupied KS levels.
In  $\varepsilon\dNL$, L refers to the lowest unoccupied ``molecular" orbital (LUMO) and $N$ to the fact that this is an eigenvalue of the KS system with $N$ electrons;
and
\begin{equation}\label{DXC1}
\Delta\dxcN = \lim\limits_{\delta N \to 0^+} \left \lbrace \left. v\dxc(\br)\right| \dNpd - \left. v\dxc(\br)\right|\dNmd \right\rbrace
\end{equation}
 is an exchange-correlation (xc) contribution that can be obtained from the xc-potential $v\dxc(\br)$ for ensemble particle densities.
Thus, $\Delta\dxcN$ is due to the discontinuities of  $ v\dxc(\br)$ that can occur
at integer electron numbers~\cite{PPLB82,GT14}. 

A few  notes should be briefly mentioned:
(a) Eq.~(\ref{FG3}) is derived by borrowing the expression of the Hartree energy from regular DFT [see Eq.~(\ref{EH-eq}) below] by evaluating it on the ensemble particle density. The result is a smooth functional of $N$ and, thus, the Hartree potential does not  contribute to the fundamental gap.
But generalizations of the Hartree-xc energy may also allow  `Hartree-like' contributions, with formal and practical advantages~\cite{GD13, KK13, KK14}.
In a different framework, a similar expression to Eq.~(\ref{FG3})  is derived without invoking fractional electron
numbers \cite{LZ14}. Moreover, in a recently derived framework,  ensemble densities and corresponding xc-functionals are employed to tackle optical and fundamental gaps in a unified fashion \cite{SF18}. In this work, however, we stay within the original EDFT formulation~\cite{PPLB82,PL97}.

 Finally, let us note that
Eq.~(\ref{FG2}) together with Eq.~(\ref{FG3}) and Eq.~(\ref{FGKS}) imply
\begin{equation}\label{DXC2}
\Delta\dxcN =  \varepsilon\dNaH -  \varepsilon\dNL\;.
\end{equation}
Thus, it should be apparent that $\Delta\dxcN$ yields in general  a non-vanishing contribution. Artificially confined many-electron systems, such as QDs, can exhibit $\Delta\dxcN$ of sizable magnitude \cite{RM02,CBKR07}.

Although Eqs.~(\ref{FG1}), (\ref{FG2}), and (\ref{FG3}) give access to the same 
fundamental gap (i.e., $G\dE \equiv G\de \equiv G\dD$), the procedures and corresponding computational efforts can differ substantially.
Equation~(\ref{FG1}) entails {\em three} distinct self-consistent calculations performed for $N-1$, $N$ and $N+1$, respectively. On the other hand,
Eq.~(\ref{FG2}) requires {\em two} independent self-consistent calculations performed for $N$ and $N+1$. Finally, Eq.~(\ref{FG3}) involves only one self-consistent calculation for $N$ electrons, once the limit is expressed analytically. Below, we come back to this point when discussing the x-only contribution in detail.
Next, let us briefly discuss approximate calculations.

It is well-known that the issue of getting vanishing $\Delta\dxcN$ -- when local-density approximation (LDA) or generalized-gradient approximation (GGA) is directly evaluated on the ensemble densities as in Eq.~(\ref{DXC1}) -- can be overcomed by adding many-body corrections as in the GW calculations~\cite{GSS88,AG98,ORR02,GMR06}.
Nevertheless, here we stick to computationally less expensive  DFT-based approaches.

For finite systems, it has been shown that the LDA and GGA forms may become useful if they are properly upgraded to EDFT\cite{KK13,KK14,AG15}. Here, instead, we proceed within a somewhat more traditional approach, to minimize both numerical and formal efforts.

A reason of inaccuracy ascribed to procedures based on LDA and GGA when computing fundamental gaps of atoms, molecules, and their arrays through Eq.~(\ref{FG2}), has been the over-damped tail of the xc-potential, which does not bind the outer electrons sufficiently (if at all). 
{\em Non} Coulombic  (e.g., harmonic) potentials
can model effectively the confinement of electrons in artificial nanostructures (such as semiconductor interfaces). When such confinements are sufficiently strong, the over-damped tail of the LDA or GGA xc-potentials may not have dramatic implications. Indeed, Capelle {\it et al.}~\cite{CBKR07} have demonstrated that LDA calculations of fundamental gaps based on Eq.~(\ref{FG2}) are equally accurate as those obtained from Eq.~(\ref{FG1}). In the same work, excellent agreement between  LDA and full configuration interaction results\cite{Rontani2006} was also pointed out. We discuss these cases in more detail below.

For the calculation of the fundamental gaps, meta-GGAs (MGGAs) are promising alternatives but still with mixed results\cite{NV11,EH14,ZHSP16}.
A class of models for the xc-potential (GGA-like and MGGA-like) have stimulated a surge of attention~\cite{GLLB95,BJ06,TB09,KOER10,H13,AK13,COM14}. Due to their 
computational simplicity and reasonable accuracy, they may offer a suitable trade-off especially in (pre-)screening of large data sets\cite{C1EE02717D}.

Reaching a satisfactory accuracy in the calculation of fundamental gaps usually requires orbital-dependent functionals, e.g., in the form of hybrids. In this case, the {\it generalized} rather than the regular KS approach is adopted as a convenient 
computational procedure, and a part of $\Delta\dxcN$ is absorbed in the corresponding {\it generalized} KS gap\cite{HSE03,EHSE,SEKB10,KSRB12,Franchini14,Periodic17}. However, hybrid-based calculations can be rather expensive computationally.

In this work,  we show that accurate estimations of the fundamental gap for QDs can be obtained by means of a computationally straightforward procedure, which requires a {\it single} set of self-consistent calculations supplied with a {\em non} self-consistent calculation of negligible computational burden -- all at the LDA level.
Our attention  was drawn to such a procedure by earlier works~\cite{CC13,EJB17} that have considered atoms, molecules, and extended systems. Here, our focus is on two-dimensional QDs -- for which, we will also analyze the case of x-only approximations extensively.

This paper is organized as follows. Theoretical preliminaries illustrating the approach and the necessary computational steps are given in Sec.~\ref{SecTheo}. 
Results of the applications are reported in Sec.~\ref{SecApp}. The paper is summarized with an outlook in Sec.~\ref{SecCon}.

\section{Theory}\label{SecTheo}

In the following, as in the typical calculations reported in the literature for QDs, we work within a spin-unrestricted formulation.  Furthermore, we focus on electrons which are effectively confined to two-spatial dimensions, which is the case of main interest when considering semiconductor QDs.~\cite{RM02}
In spin-DFT~\cite{vBH72} (SDFT), under the restriction of collinear spin polarization, the total energy, $E$, of 
$N$ interacting electrons in a given (local) external potential (i.e., the confinement), $v_{0\sigma}(\vr)$,
can be expressed as functional of the two spin densities $n_{\sigma}(\vr)$ (with  $\sigma=\uparrow,\downarrow$)
\bea
E[\un] &=&  T\dks[\un]  
+ E\dHartree[n] + E\dxc[\un] \nn \\
&+& \sum_{\sigma = \uparrow,\downarrow} \int d\uD r  \; v_{0\sigma}(\vr) n_{ \sigma }(\vr)\,,
\label{etot-sdft}
\eea
where $ d\uD r$ is the infinitesimal volume in two dimensions, $\vr = (x, y)$ is the position vector and  $x$ and $y$ are the coordinates, $\un$ denotes the pair $(n_{\uparrow}, n_{\downarrow})$, 
$n = n_{\uparrow} + n_{\downarrow}$ 
is the total particle density. 
$T\dks[\un]$ is the kinetic energy of the Kohn-Sham systems,
which is defined as
\be
T\dks[\un] = 
\sum_{\sigma = \uparrow,\downarrow} \sum_{j=1}^{N_\sigma}
\int d\uD r  \; \varphi\djs^*(\vr) \left( - \frac{\nabla^2}{2} \right) 
\varphi\djs(\vr) \,;
\label{ts}
\ee
here the Lapalcian takes into account only two-dimensional partial derivatives, namely $\nabla^2 =  \partial^2_x + \partial^2_y$.
$N_\sigma$ is the number of electrons with spin $\sigma$, and $N=N_{\uparrow}+N_{\downarrow}$.
$E\dHartree[n]$ is the (Hartree) electrostatic interaction energy defined as
\be\label{EH-eq}
E\dHartree[n] = \frac{1}{2} \int d\uD r  \int d\uD r' \; \frac{n(\vr) n(\vr')}{| \vr - \vr'|} \, .
\ee
Finally, $E\xc[\un]$ is the exchange-correlation energy functional that in practice needs to be approximated. 

The KS single-particle orbitals are solutions of the equations~\cite{vBH72} 
\be 
 -\frac{\nabla^2}{2} \varphi\djs(\vr)  + v\dkss[\un](\vr) \varphi\djs(\vr) 
= \varepsilon\djs \varphi\djs(\vr) \,.
\label{kseq}
\ee
The KS potential may be decomposed as
\be
v\dkss[\un](\vr) = v_{0\sigma}(\vr) + v\dHartree[\un](\vr) + v{\dxcs}[\un](\vr) \,,
\label{espp}
\ee
where
\be
v\dHartree(\vr)[n] = \int d\uD r' \frac{n(\vr')}{| \vr - \vr'|} \,,
\ee
and
\be
v\dxcs[\un](\vr) = \frac{\delta E\dxc[\un]}
{\delta n_{\sigma}(\vr)} \,.
\label{fdexc}
\ee
The exact spin-densities can be calculated from the exact KS orbitals, in principle, by summing $n\djs(\vr) =| \varphi\djs(\vr) |^2 $ over the occupied single-particle states, $n_{\sigma}(\vr) = 
\sum_{j=1}^{N_\sigma} n\djs(\vr)$.

As mentioned in the introduction, the KS scheme provides us with all the ingredients to compute the fundamental gap either via differences of total energies [as in Eq.~(\ref{FG1}) ] or KS eigenvalues [as in Eq.~(\ref{FG2})]. 
In next subsection, however, we are after the third (approximate) procedure, which is suggested by working with Eq.~(\ref{FG3}) at the level of the exchange-only approximation.

\subsection{From exact to approximate x-only expressions}\label{SubKey}

 Ensemble-SDFT 
allows us to consider a fractional number of 
electrons, which are realized by mixing pure states with 
different integer numbers of electrons. The  
ensemble xc-potential can jump by a well-defined 
(spin-dependent) constant, whenever the number of electrons 
passes through an integer value. This leads to an 
appealing way to compute the fundamental gap~\cite{PPLB82} [see 
Eq.~(\ref{FG3})]. 

To conclude our analysis, however, we do not go into the details of ensemble-SDFT. It is sufficient to recall that through Eq.~(\ref{DXC1}) we can isolate the exact x-contribution to the fundamental gap as follows \cite{Perdew85,PhysRevA.52.4493}:
\bea\label{DDKLI} 
\Delta \dxN = \langle u_{\scriptscriptstyle\rm 
XL\sigma}[\un] \rangle_{\scriptscriptstyle\rm L\sigma} - 
\langle v\dxs[\un]\rangle_{\scriptscriptstyle\rm 
L\sigma}\,.
\eea
where 
\begin{align}\label{uxsi}
u\dxjs[n](\vr) =& - \sum_{i=1}^{N_\sigma} \frac{ \varphi^*\dis(\vr) }{ \varphi\djs^*(\vr) }
  \int d\uD r' \frac{\varphi^*_{j\sigma}(\vr') \varphi\dis(\vr')}{|\vr -\vr'|} \,,
\end{align}
\begin{align}\label{vbar}
\langle  v\dxs[\un]  \rangle\djs  = 
  \int d\uD r  \; 
\varphi\djs^*(\vr) v\dxs[\un](\vr) \varphi\djs(\vr) \,,
\end{align}
and
\begin{align}\label{ubar}
\langle  u\dxjs [\un]  \rangle\djs = \int d\uD r  \, 
\varphi\djs^*(\vr) u\dxjs[\un](\vr) \varphi\djs(\vr) \,.
\end{align}
For later convenience,  we emphasize that the above quantities are well defined also for  ${j\sigma} \ne {\rm L\sigma}$.
Writing Eq.~(\ref{DDKLI}), we have assumed that the variation of the electron number occurs only within a given spin channel. For the sake of simplicity, we have also assumed that the considered states do not involve degeneracies.

So far, exchange and correlation were included and treated exactly. Next, we neglect the correlation and restrict ourselves to the exact-exchange-only approximation (EXX). Thus
\begin{widetext}
\be
E\dxc \rightarrow E\dx = -\frac{1}{2} \sum_{\sigma=\uparrow,\downarrow} \sum_{i,k=1}^{N_\sigma} 
  \int d\uD r' \int d\uD r'' \frac{ \varphi_{i \sigma}(\vr'') \varphi_{i \sigma}^*(\vr') \varphi_{k \sigma}^*(\vr'') \varphi_{k \sigma}(\vr')}{|\vr'
- \vr''|} \;.
\label{Ex}
\ee
\end{widetext}
First we notice that $E\dx$ depends on $\un$ implicitly, i.e., through the KS orbitals $\varphi\djs(\vr) \equiv \varphi\djs[\un](\vr) $. Thus, in the case of Eq.~(\ref{Ex}), the evaluation of the functional derivative as in Eq.~(\ref{fdexc})
requires the solution of an integral equation for the EXX 
potential, to be used self-consistently in the solution of the KS equations~\cite{SH53,TS76, GRABO1995141, Grabo1997, KK08}.
In what follows, however, we simplify both our numerical efforts and analysis by adopting the Krieger, Li and Iafrate (KLI) approximation~\cite{KLI90a,KLI92}.

The EXX potential in the KLI approximation is given by
\bea\label{vxKLI}
v\uKLI\dxs[\un](\vr) &=& v\dSLs[\un](\vr)  + \Delta v\uKLI\dxs[\un](\vr),
\eea
where
\be\label{SL}
v\dSLs[\un](\vr) = \frac{1}{ n_{\sigma}(\vr)}\sum_{j=1}^{N_\sigma}   n\djs(\br)  u\dxjs[\un](\vr)
\ee
is the Slater (SL) potential and
\begin{multline}\label{DKLI}
 \Delta v\uKLI\dxs[\un](\vr) = \frac{1}{n_{\sigma}(\vr)}\sum_{j=1}^{N_\sigma}   
 n\djs(\br)  \\
 \times \left[ \langle  v\uKLI\dxs[\un] \rangle\djs  -  \langle  u\dxjs[\un]  \rangle\djs  \right]
\end{multline}
can be regarded as a correction to the Slater potential. 

As long as the particle density and the spin-polarization are preserved,
the KS potential can be shifted, for each spin channel, by an arbitrary constant and thus the term with $j=N_{\sigma}$ in Eq.~(\ref{DKLI}) can be set to zero.
It may also be useful  to remind that for strongly confined systems such as QDs -- which are the systems of  interest in this work --, the Slater potential yields the leading contribution to the x-only potential and vanishes for $r \to +\infty$ (Refs.~\onlinecite{PRHG07,RPPG09}).


 Next, we seek to further minimize our numerical efforts. 
As shown in Appendix~\ref{A1},  elementary but tedious algebraic steps allow us to
define an {\em approximation} to $\Delta\dxN $ in terms of the difference of single-particle energies, as follows
\bea\label{keyKLI}
\Delta\dxN\uKLI = \widetilde{\varepsilon}\uKLI \dNaHs   - \varepsilon\uKLI \dNLs \,.
\eea
In Eq.~(\ref{keyKLI}), $\widetilde{\varepsilon}\uKLI \dNaHs$ is a single-particle energy that refers to the system with $N+1$ electrons but it is obtained 
by
using as an input  the single-particle orbitals from the (self-consistent solution of) the corresponding $N$-electron problem -- hence, the
 {\em tilde} is used here  to stress  that   ``frozen'' orbitals are employed. 
$\widetilde{\varepsilon}\uKLI \dNaHs$ can be computed  through a single iteration of the EXX-KLI procedure. In this step,
the KS potential must be shifted -- at most by a constant value -- such that it goes to zero at large distance from the system. 
Thus,  $\widetilde{\varepsilon}\uKLI \dNaHs$  may be  related to an approximate ionization potential for the systems with $N+1$ electrons.
 $\varepsilon\uKLI \dNLs$ is obtained as usual from the self-consistent solution for the system with $N$-electrons.

The importance of Eq.~(\ref{keyKLI}) is in the fact that it readily suggests us that a {\em non-vanishing} -- albeit approximate --  ${\Delta}\dxN$ may be obtained by
replacing the EXX-KLI quantities 
with quantities that do not necessarily entail orbital-dependent functionals.
Especially for the systems considered in this work, it is compelling to try with the simplest approximation
\bea\label{keyXLDA}
 {\Delta}\dxN\uKLI \rightarrow {\Delta}\dxN\uLDA  := \widetilde{\varepsilon}\uxLDA \dNaHs   - \varepsilon\uxLDA \dNLs \,,
\eea
where LDA, for brevity, stands for local-spin-density approximation, and
the notation emphasizes that eigenvalues are determined within {\em x-only} LDA calculations.
Eq.~(\ref{keyXLDA}) requires no extra implementations, in codes that already implement regular calculations (including a restart procedure from given orbitals and the control of the number of iterations). 
Further details on the numerical procedure are reported in the section devoted to our applications (see below). \\

\subsection{Inclusion of correlation } 
It is tempting to extend Eq.~(\ref{keyXLDA}) to include the  correlation as follows:
\begin{align}\label{keyXCLDA}
{\Delta}\uLDA\dxcN := \widetilde{\varepsilon}\uLDA\dNaHs - \varepsilon\uLDA\dNLs\;.
\end{align}
This equation expressed through the xc-potential [see Eq.~(\ref{step5})] 
has been previously suggested in Ref.~\onlinecite{CC13} and --
with improved models for the xc-potential~\cite{GLLB95,KOER10} -- also in Ref.~\onlinecite{EJB17}.
Comparing Eq.~(\ref{keyXCLDA}) with Eq.~(\ref{DXC2}), we see that
Eq.~(\ref{keyXCLDA}) not only invokes an `LDA replacement' but  
also makes use frozen orbitals [similarly as in Eq.~(\ref{keyXLDA})].
In Ref.~\onlinecite{CC13} it is shown that
$\widetilde{\varepsilon}\dNaHs$ can be connected to ${\varepsilon}\dNaHs$ in a perturbative fashion -- but we will not explore such corrections in this work.
In Refs.~\onlinecite{CC13,EJB17} neither electrons in artificial confinements nor the x-only limit were scrutinized. We carry out these analyses on QDs in the next section.

\section{Applications}\label{SecApp}

In this section, we show  that the fundamental gaps of QDs computed up to exchange-only effects by using Eq.~(\ref{keyXLDA}) compare very well with those obtained by using Eq.~(\ref{keyKLI}).
More importantly, we show that the estimations  including correlations through Eq.~(\ref{keyXCLDA}) are notable as well.

\subsection{Quantum-dot model and numerical methods}
We model electrons in a semiconductor QD with a two-dimensional harmonic external potential in effective atomic units~\cite{effau} as
\begin{equation}\label{EqEllPot}
v_{0{ \sigma}}(\vr) = \frac{1}{2}\omega^2(x^2+\alpha^2y^2) \;,
\end{equation}
where $\omega$ determines the strength of the confinement, and $\alpha$ defines the elliptical 
deformation.
 The harmonic confinement is the standard approximation for electrons
in semiconductor QDs.~\cite{RM02}  We use the material parameters of GaAs, $m^*=0.067 m_e$ and $\epsilon=12.4 \epsilon_0$.
In practice, the purpose of the ellipticity is to model more realistic QDs that
are not perfectly symmetric due to deformations and impurities, etc.
For the x-only calculations in Sec.~\ref{xresults}, we set $\alpha=1.05$ corresponding to an eccentricity 
of $e\approx 0.30$. These cases are free from degeneracies of the relevant single-particle levels. 
Whereas in Sec.~\ref{xcresults}, we set $\alpha=1$ to compare with  numerically exact results for conventional parabolic QDs -- some of these cases include degeneracies.
In all the cases, however, we could employ integer occupation numbers.

We carry out all our calculations with the OCTOPUS code~\cite{octopus15,octopus06,octopus03}, that solves the KS equations on a regular grid with Dirichlet boundary conditions.

We select a grid spacing of $g=0.1/\sqrt{\omega}$ eff. a.u.
The simulation box containing the real-space domain is circular with a radius 
of $R=K/\sqrt{\omega}$, where $K=5.0$ eff. a.u. is used for $N=2,4,5$ and $K=\lbrace 6.0, 6.5, 7.0, 7.5, 8.0, 8.5\rbrace$ eff. a.u. is used for $N=\lbrace 6,12,20,30,42,56 \rbrace$, respectively. The self-consistent criteria for the solution of the KS equation is $\epsilon = \int d\vr \left[ n^{old}(\vr)-n^{new}(\vr) \right]/N < 10^{-6}$. 
We verified numerically that these parameters are sufficient to get fundamental gaps converged within the fourth significant digit.

\subsection{Exchange-only results}\label{xresults}

\begin{figure}
\begin{center}
\includegraphics[width=1.0 \linewidth,=1]{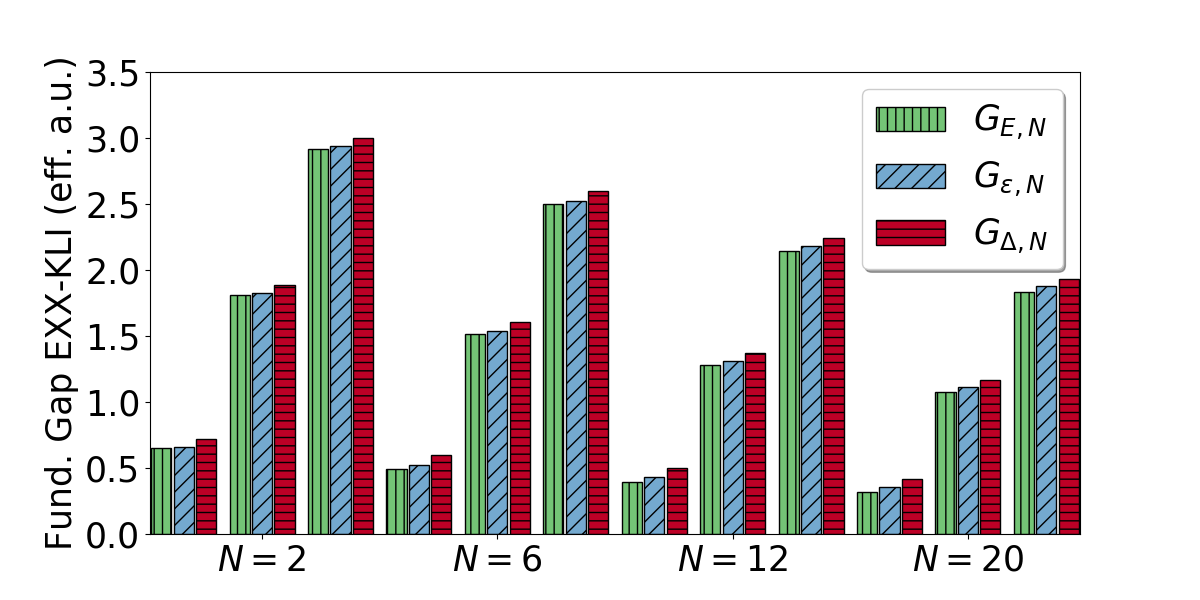}
\caption{
EXX-KLI results for the fundamental gaps computed according to  $G\dE$ [Eq. (\ref{FG1})],  $G\de$ [Eq. (\ref{FG2})], and $G\dD$ [Eq. (\ref{FG3}) together with Eq. (\ref{keyKLI})]. For each $N$, the bars from left to right correspond to $\omega = 0.50$, $1.50$, and $2.50$, respectively and $\alpha=1.05$ [see Eq.~(\ref{EqEllPot})].}
\label{fig1}
\end{center}
\vskip -0.5cm
\end{figure}

In Fig.~\ref{fig1} we show the fundamental gaps resulting from our EXX-KLI calculations for QDs with 
$N=2,\ldots,20$ electrons.
The considered confinements are such $\alpha=1.05$ and
$\omega = 0.50$, $1.50$, and $2.50$, corresponding to
the three sets of bars for each $N$ in Fig.~\ref{fig1}, respectively. We compare 
the results for the EXX-KLI fundamental gap obtained 
by means of three different procedures as suggested by Eqs.~(\ref{FG1}), (\ref{FG2}), and (\ref{FG3}). 
According to Fig.~\ref{fig1}, the values for the gaps given by the aforementioned expressions are relatively close to each other in all cases. 
We stress that no deviations would be observed if the exact xc-energy functional could be used.
These results support in particular the usefulness of Eq. (\ref{FG3}), which corresponds to the simplest procedure [see also Eq.~(\ref{keyKLI})].

\begin{figure}
\begin{center}
\includegraphics[width=1.0 \linewidth,=1]{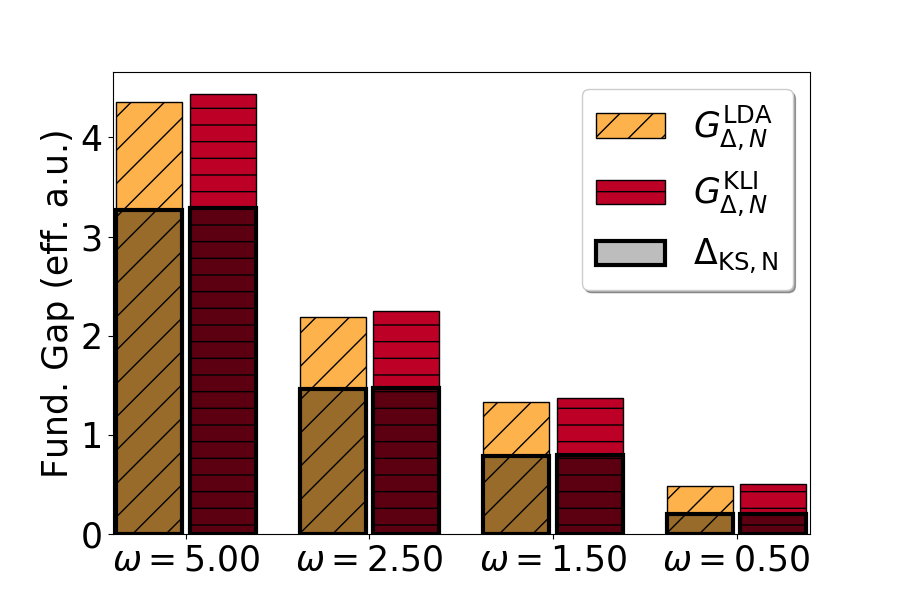}
\caption{
Fundamental gaps $G_{\Delta,N}$ obtained with the exchange-only KLI and LDA approximations, respectively, for elliptic quantum dots [Eq.~(\ref{EqEllPot}) with $\alpha=1.05$]
with $N=12$ electrons and varying confinement strength $\omega$. The contributions of the Kohn-Sham gap [Eq.~(\ref{FGKS})] are
marked by shaded open boxes. The remaining part is given by the derivative discontinuity, that is, Eq.~(\ref{keyKLI}) and
Eq.~(\ref{keyXLDA}) in the case of KLI and LDA, respectively. All the numerical results are given in Table~\ref{table1}.
}\label{fig2}
\end{center}
\vskip -0.5cm
\end{figure}

\begin{figure}
\begin{center}
\includegraphics[width=1.0 \linewidth,=1]{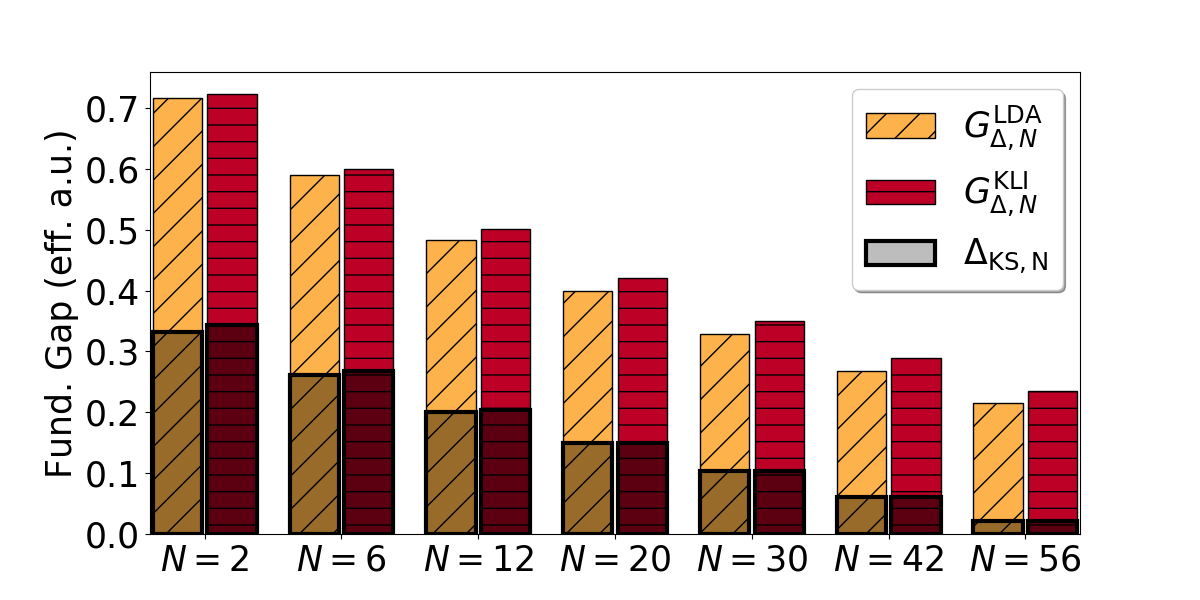}
\caption{Same as Fig.~(\ref{fig2}) but for a fixed value of the confinement strength $\omega=0.5$ and varying number
of electrons $N$. 
}\label{fig3}
\end{center}
\vskip -0.5cm
\end{figure}

Next we compare our EXX-KLI results based on Eq.~(\ref{keyKLI}) with the simpler and numerically more efficient LDA calculations as performed according to Eq.~(\ref{keyXLDA}).
The results are reported in Table~\ref{table1} in the Appendix. Some of the key results are visualized 
for fixed $N=12$ and variable $\omega$ in Fig.~\ref{fig2}, 
and for fixed $\omega=0.5$ and variable $N$ in Fig.~\ref{fig3}. Generally, the LDA values computed according to 
Eqs.~(\ref{FGKS}) and (\ref{keyXLDA}) agree well with the EXX-KLI approximation:
the mean absolute relative deviations being only $4\%$, with
a maximum deviation of $8\%$.
The LDA errors in the fundamental gap are mostly due to the derivative discontinuity. This can be seen in the KS gaps (open boxes in Figs.~\ref{fig2} and \ref{fig3}) that are in most cases very close to each other. Equation~(\ref{keyXLDA}) underestimates the EXX-KLI 
discontinuity but {\it only slightly} in most cases. 

\subsection{Results including correlations}\label{xcresults}

\begin{figure}
\begin{center}
\includegraphics[width=1.1 \linewidth,=1]{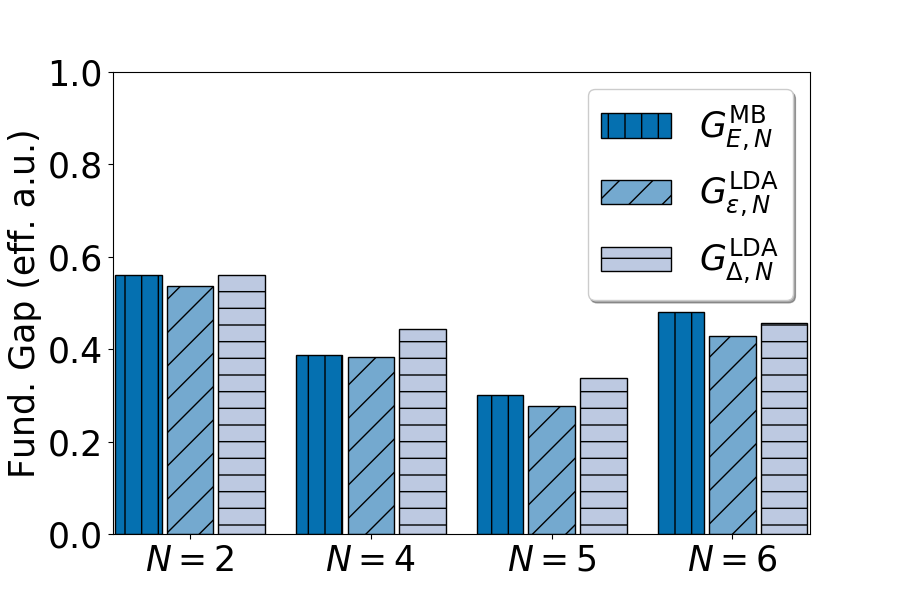}
\caption{Fundamental gaps including correlations for parabolic quantum dots [Eq.~(\ref{EqEllPot}) with $\alpha=1$] with a fixed confinement 
strength of $\omega=0.35$ and variable number of electrons $N$.  $G_{E,N}^{\rm MB}$ is the full configuration interaction value from Ref.~\onlinecite{CBKR07}; $G_{\epsilon,N}^{\rm LDA}$ is obtained from Eq. (\ref{FG2}) at the LDA level; $G_{\Delta,N}^{\rm LDA}$ from Eq. (\ref{keyXCLDA}).
See also Table~\ref{table2}.
}\label{fig4}
\end{center}
\vskip -0.5cm
\end{figure}

Finally, we consider the full gaps when including correlations. We consider parabolic QDs by setting $\alpha = 1$ in Eq.~(\ref{EqEllPot}) and compare our results against exact diagonalization 
results reported in Ref.~\onlinecite{CBKR07}. 
Although
alternative methodologies to direct exact diagonalization have been developed,~\cite{YNPRBH17} 
large benchmark data sets are still challenging to be produced.

Fig.~\ref{fig4} shows the results for
$\omega=0.35$ and $N = 2 \ldots 6$. All the values  -- along with additional cases for different $\omega$ -- can be found in Table~\ref{table2} of the Appendix. 
Since
the values of the exact KS gaps are not available, KS gaps are not highlighted.
The agreement between our scheme and the many-body (MB) results is reasonable with a mean absolute error of $14\%$. 

We stress that our procedure exploits Eq.~(\ref{keyXCLDA})
as in 
$G\uLDA\dD =\Delta\uLDA\dksN+\Delta\uLDA\dxcN = \widetilde{\varepsilon}\uLDA\dNaHs  -\varepsilon\uLDA\dNHs $,
while the LDA procedure of Ref.~\onlinecite{CBKR07} -- for which data is also shown both Fig.~(\ref{fig4}) and in 
Table~\ref{table2} of the Appendix -- computes
$G\uLDA\de  = \varepsilon\uLDA\dNaHs-\varepsilon\uLDA\dNHs$.
Thus when comparing $G\uLDA\dD$ with $G\uLDA\de$,  the systematic overestimation $G\uLDA\dD \ge G\uLDA\de$ may be explained in terms of the lack of relaxation of the frozen orbitals which are used in Eq.~(\ref{keyXCLDA}).

\section{Conclusions and outlook}\label{SecCon}
In this work, we have given evidence that the fundamental gaps of artificially confined systems such as semiconductor quantum dots can be accurately estimated by means of a simple procedure within a minimal computational effort: a regular Kohn-Sham
calculation plus a straightforward non-self-consistent (one-shot) evaluation -- all carried within the local-density approximation. Specifically, we have considered the case of quantum dots defined by parabolic and elliptical confinements.

It would be interesting to explore whether our conclusions can apply also to a larger variety of artificially confined nanoscale systems. Corrections in the form of the gradients of the particle-density may help to preserve accuracy without substantially increasing the numerical effort. But functional forms that explicitly depend only on the particle density and, possibly, gradients thereof, 
 can still  fail in the case of periodic systems~\cite{EJB17} for which, an approach based on forms considered in Refs.~\onlinecite{GLLB95,KOER10,EJB17} (if properly extended also to lower dimensions) appears to be the most promising.

\section*{Acknowledgments}
The authors thank Alice Ruini and Tim Gould for useful discussions.
C.A.R and S.P. acknowledge support from the European Community through the FP7's Marie-Curie
International-Incoming Fellowship, Grant agreement No.
623413.

\appendix

\section{Derivation of Equation (\ref{keyKLI})}\label{A1}

 Let us start with the self-consistent EXX-KLI solution of a closed-shell $N$-electron system. 
As before, we assume non degeneracy  for  the relevant occupied and {\em non} occupied single-particle levels (within each spin channel).

Next, let us add one electron to the system and keep the single-particle orbitals  {\it frozen}; i.e., equal to
the orbitals of the $N$-electron system. Let  the `additional' electron be in the spin channel $\sigma$.
The spin density for the $(N+1)$-electron system is, thus, given by $\widetilde{n}\ds = n\ds + | \widetilde{\varphi}\dNaHs|^2$, where  $\widetilde{\varphi}\dNaHs  \equiv \varphi\dNLs$ and $n\ds$  is the spin-density of the $N$-electron system. No modification needs to be considered in the {\em other} spin channel.
The corresponding x-potential, $v\dxs[\widetilde{\un}]$, can be readily expressed in the EXX-KLI approximation [see
Subsection~\ref{SubKey}].  We remind that $v\dxs[\widetilde{\un}]$ may be shifted  by a constant in such a way
\begin{equation}\label{step1}
\langle v\dxs[\widetilde{\un}]\rangle_{\scriptscriptstyle\rm H\sigma} -  \langle u_{\scriptscriptstyle\rm XH\sigma}[\widetilde{\un}] \rangle_{\scriptscriptstyle\rm H\sigma} \equiv  0\,.
\end{equation}

Now, let us consider the single-particle energies
\be\label{step2}
\widetilde{\varepsilon}\uKLI\dNaHs  = \langle  \hat{h}\dzs + v\dHartree[\widetilde{n}] +  v\uKLI\dxs[\widetilde{\un}] \rangle_{\scriptscriptstyle\rm H\sigma}
\ee
for the HOMO of the system with $N+1$ electrons, and
\begin{align}\label{step3}
\varepsilon\uKLI\dNLs  
=  \langle  \hat{h}\dzs + v\dHartree[{n}] +  v\dxs\uKLI[{\un}] \rangle_{\scriptscriptstyle\rm L\sigma} \,
\end{align}
for the LUMO of the system with $N$ electrons.
Note that
$ \hat{h}\dzs(\br) = - \nabla^2 /2+ v\dzs(\br)$ 
and $v\dHartree[\widetilde{n}] = v\dHartree[n] +  v\dHartree[| \widetilde{\varphi}\dNaHs|^2]$. 
Thus
the difference of Eq.~(\ref{step2}) and Eq.~(\ref{step3}) can be readily written as follows
\begin{eqnarray}
\label{step5}
 \widetilde{\varepsilon}\dNaHs\uKLI   - \varepsilon\dNLs\uKLI
& = &  \langle v\dxs\uKLI[\widetilde{\un}]\rangle_{\scriptscriptstyle\rm H\sigma}  - \langle v\dxs\uKLI[\un]\rangle_{\scriptscriptstyle\rm L\sigma}  \nn \\ &+& \langle v\dHartree[| {\varphi}\dNLs|^2]   \rangle_{\scriptscriptstyle\rm L\sigma}.
\end{eqnarray} 
Next, Eq.~(\ref{step1}) together with the identity
\begin{equation}
    \langle u_{\scriptscriptstyle\rm XH\sigma}[\widetilde{\un}] \rangle_{\scriptscriptstyle\rm H\sigma} \equiv  \langle u_{\scriptscriptstyle\rm XL\sigma}[\un] \rangle_{\scriptscriptstyle\rm L\sigma} - \langle v\dHartree[| {\varphi}\dNLs|^2] \rangle_{\scriptscriptstyle\rm L\sigma}
\end{equation}
allow us to rewrite Eq.~(\ref{step5}) as follows
\begin{eqnarray}\label{step6}
 \widetilde{\varepsilon}\dNaHs\uKLI   - \varepsilon\dNLs\uKLI
& = &  \langle u_{\scriptscriptstyle\rm XL\sigma}[\un] \rangle_{\scriptscriptstyle\rm L\sigma} - \langle v\dxs\uKLI[\un]\rangle_{\scriptscriptstyle\rm L\sigma}\,.
\end{eqnarray}
Note that in the steps above, we have repeatedly used
$\widetilde{\varphi}\dNaHs  \equiv \varphi\dNLs$.

Evaluating  Eq.~(\ref{DDKLI})  on EXX-KLI quantities and comparing with Eq.~(\ref{step6}), we conclude that
\begin{eqnarray}\label{akeyKLI}
\Delta\dxN\uKLI   \equiv \widetilde{\varepsilon}\uKLI \dNaHs   - \varepsilon\uKLI \dNLs   \,.
\end{eqnarray}
Note, the KLI approximation is not essential -- it is used here for simplicity. Correlation forms restricted to have an explicit dependence only on occupied orbitals may also be easily accommodated.

\section{Tables of the numerical results}

\begin{table}[h]
\begin{center}
\begin{tabular}{|c|c|c|c|C{1.5cm}|C{1.5cm}|c|c|}
\hline
~& ~&\multicolumn{2}{|c|}{~$\Delta\dks$~} & \multicolumn{2}{|c|}{~$\Delta\dx := \widetilde{\varepsilon}\dNaH - \varepsilon\dNL$~} & \multicolumn{2}{|c|}{~$\Delta\dks +\Delta\dx$~}\\
\hline
\hline
$\omega$ & $N$ & ${\mathrm{LDA}}$ & ${\mathrm{KLI}}$ & ${\mathrm{LDA}}$ & ${\mathrm{KLI}}$ & ${\mathrm{LDA}}$ & ${\mathrm{KLI}}$\\
\hline
$5.00$	&	$2$  &	$4.31$	&	$4.37$	&	$1.30$	&	$1.33$	&	$5.61$	&	$5.70$\\
$5.00$	&	$6$	&	$3.77$	&	$3.81$	&	$1.19$	&	$1.23$	&	$4.96$	&	$5.03$\\
$5.00$	&	$12$	 &	$3.27$	&	$3.29$	&	$1.09$	&	$1.15$	&	$4.36$	&	$4.44$\\
$5.00$	&	$20$	 &	$2.82$	&	$2.82$	&	$0.99$	&	$1.07$	&	$3.80$	&	$3.90$\\
$5.00$	&	$30$ &	$2.38$	&	$2.38$	&	$0.90$	&	$1.01$	&	$3.28$	&	$3.39$\\
$5.00$	&	$42$	 &	$1.95$	&	$1.95$	&	$0.84$	&	$0.95$	&	$2.79$	&	$2.90$\\
$5.00$	&	$56$ &	$1.54$	&	$1.53$	&	$0.79$	&	$0.90$	&	$2.32$	&	$2.43$\\
\hline	
\hline							
$2.50$	&	$2$ &	$2.04$	&	$2.08$	&	$0.91$	&	$0.92$	&	$2.95$	&	$3.00$\\
$2.50$	&	$6$ &	$1.73$	&	$1.76$	&	$0.82$	&	$0.84$	&	$2.55$	&	$2.59$\\
$2.50$	&	$12$	&	$1.46$	&	$1.47$	&	$0.73$	&	$0.77$	&	$2.19$	&	$2.25$\\
$2.50$	&	$20$	&	$1.21$	&	$1.22$	&	$0.66$	&	$0.72$	&	$1.87$	&	$1.93$\\
$2.50$	&	$30$	&	$0.98$	&	$0.98$	&	$0.60$	&	$0.67$	&	$1.58$	&	$1.65$\\
$2.50$	&	$42$	&	$0.75$	&	$0.75$	&	$0.55$	&	$0.63$	&	$1.31$	&	$1.38$\\
$2.50$	&	$56$	&	$0.54$	&	$0.53$	&	$0.52$	&	$0.59$	&	$1.06$	&	$1.13$\\
\hline
\hline							
$1.50$	&	$2$		&	$1.16$	&	$1.19$	&	$0.69$	&	$0.70$	&	$1.85$	&	$1.89$\\
$1.50$	&	$6$		&	$0.97$	&	$0.98$	&	$0.62$	&	$0.63$	&	$1.58$	&	$1.61$\\
$1.50$	&	$12$	&	$0.79$	&	$0.80$	&	$0.54$	&	$0.58$	&	$1.33$	&	$1.37$\\
$1.50$	&	$20$	&	$0.64$	&	$0.64$	&	$0.48$	&	$0.53$	&	$1.12$	&	$1.17$\\
$1.50$	&	$30$	&	$0.49$	&	$0.49$	&	$0.44$	&	$0.49$	&	$0.93$	&	$0.98$\\
$1.50$	&	$42$	&	$0.36$	&	$0.35$	&	$0.41$	&	$0.46$	&	$0.76$	&	$0.81$\\
$1.50$	&	$56$	&	$0.23$	&	$0.23$	&	$0.38$	&	$0.43$	&	$0.61$	&	$0.65$\\
\hline
\hline								
$0.50$	&	$2$		&	$0.33$	&	$0.34$	&	$0.38$	&	$0.38$	&	$0.72$	&	$0.72$\\
$0.50$	&	$6$		&	$0.26$	&	$0.27$	&	$0.33$	&	$0.33$	&	$0.59$	&	$0.60$\\
$0.50$	&	$12$	&	$0.20$	&	$0.20$	&	$0.28$	&	$0.30$	&	$0.48$	&	$0.50$\\
$0.50$	&	$20$	&	$0.15$	&	$0.15$	&	$0.25$	&	$0.27$	&	$0.40$	&	$0.42$\\
$0.50$	&	$30$	&	$0.10$	&	$0.10$	&	$0.23$	&	$0.25$	&	$0.33$	&	$0.35$\\
$0.50$	&	$42$	&	$0.06$	&	$0.06$	&	$0.21$	&	$0.23$	&	$0.27$	&	$0.29$\\
$0.50$	&	$56$	&	$0.02$	&	$0.02$	&	$0.19$	&	$0.21$	&	$0.21$	&	$0.23$\\
\hline
\end{tabular}
\end{center}
\caption{ 
Fundamental gaps of elliptic quantum dots [Eq.~(\ref{EqEllPot}) with $\alpha=1.05$]  are reported together with the 
the contributions of the corresponding Kohn-Sham (KS) gap and exchange-only (x) discontinuities within two procedure that employ either the KLI or the local-density approximation. For the x-discontinuities, the KLI calculations use Eq.~(\ref{keyKLI}) while the LDA calculations use Eq.~(\ref{keyXLDA}). Values in effective atomic units\cite{effau}.
}\label{table1}
\end{table}

\begin{table}[h]
\begin{center}
\begin{tabular}{|c|c|c|c|c|}
\hline
$N$ & $\omega$  & $G_{\Delta,N}^{\rm LDA}$ &    $G_{\varepsilon,N}^{\rm LDA}$ & $G^{\rm MB}_{E,N}$ \\
\hline
\hline
$2$	&	$0.35$	&	$0.56$	&	$0.53$	&	$0.56$\\
\hline	
\hline		
$4$	&	$0.15$	&	$0.26$	&	$0.22$	&	$0.22$\\
$4$	&	$0.25$	&	$0.36$	&	$0.31$	&	$0.32$\\
$4$	&	$0.35$	&	$0.44$	&	$0.38$	&	$0.39$\\
\hline
\hline					
$5$	&	$0.15$	&	$0.21$	&	$0.17$	&	$0.20$\\
$5$	&	$0.25$	&	$0.28$	&	$0.23$	&	$0.24$\\
$5$	&	$0.35$	&	$0.34$	&	$0.28$	&	$0.30$\\
\hline
\hline					
$6$	&	$0.15$	&	$0.23$	&	$0.21$	&	$0.25$\\
$6$	&	$0.25$	&	$0.35$	&	$0.32$	&	$0.38$\\
$6$	&	$0.35$	&	$0.46$	&	$0.43$	&	$0.48$\\
\hline
\end{tabular}
\end{center}
\caption{
Fundamental gaps of parabolic quantum dots [Eq.~(\ref{EqEllPot}) with $\alpha=1$]. $N$. $G_{\Delta,N}^{\rm LDA}$ is obtained from Eq. (\ref{keyXCLDA}); $G_{\epsilon,N}^{\rm LDA}$ from Eq. (\ref{FG2}) at the LDA level;  $G_{E,N}^{\rm MB}$ is the full configuration interaction value from Ref.~\onlinecite{CBKR07}. Values in effective atomic units\cite{effau}.
}\label{table2}
\end{table}

\end{document}